# Trajectories in Logarithmic Potentials

Erwin A. T. Wosch

**Abstract**

Trajectories in logarithmic potentials are investigated by taking as example the motion of an electron within a cylindrical capacitor. The solution of the equation of motion in plane polar coordinates, (r,φ) is attained by forming a series expansion of r and of 1/r as a function of φ. The terms of the series contain polynomials, the recurrence relation of which is given, together with some further characteristics. By the comparison-theorem of infinite series, the convergence of the solution is demonstraded. The simplest trajectories in logarithmic potentials are represented by rosette type orbits with a period of 4π/3, and by circular paths.

## 1. Introduction

The path of a point mass in a potential V(r) can be described by the differential equation:

$$\frac{d}{d\varphi}\left(\frac{1}{r^2}\frac{dr}{d\varphi}\right) - \frac{1}{r} + \left(\frac{m}{L^2}\right) r^2 \cdot \frac{dV(r)}{dr} = 0 \qquad (1)$$

in accordance with classical mechanics. It is restricted to plane movements [1, 2, 3]. L denotes the angular momentum, m the mass, and r and φ the plane polar coordinates.

Closed solutions for differential equation (1) are known only for some special potentials [1, 2, 3]. These are, for exemple, the harmonic oscillator: $V(r) \sim r^2$, and the Kepler potential: $V(r) \sim 1/r$, which exist as power functions. In many other cases equation (1) can only be solved by numerical methods. Even for the logarithmic potential $V(r) \sim \ln(r)$, only numerical and no mathematical/analytical solutions exist up to now [1-11, 18, 19].

In astronomy, the logarithmic potential is of special importance for the calculation of the orbits of single stars in the galaxy [3, 4]. A further illustrative example is to be found in the field of electronic optics, involving the determination of the electron orbit within the cylinder capacitor [12, 13]. Typical for orbits in logarthmic potentials is the determination of movement of a test charge q = -e in the cylinder capacitor. Formally this case can be treated by equation (1) by the substitution: m → q·m = -e·m.

## 2. Equation of Motion for an Electron in a Cylinder Capacitor

Apart from stray fields at the ends of the cylinder, the electric field prevailing within a cylinder capacitor is

$$\vec{E} = \frac{U_C}{\ln(R_2/R_1)} \cdot \frac{\vec{e_r}}{r} \tag{2}.$$

This electric field may be considered to be stationary. The constant $k_0$ contains the capacitor voltage $U_C$ and the inner and outer radii $R_1$ and $R_2$ (see figure 1)

$$k_0 = \frac{U_C}{\ln(R_1/R_2)} \quad , \quad k_0 < 0 \tag{3}.$$

The electric potential for this is:

$$V(r) = k_0 \cdot \ln(r/R_1) \tag{4},$$

the potential of the inner radius $R_1$, being put to zero.

The following considerations assume that the electron enters the capacitor orthogonally to the z axis, and that the orbits to be calculated lie in the x-y plane (plane orbits). The motion along the z axis is a minor problem and will not be considered here.

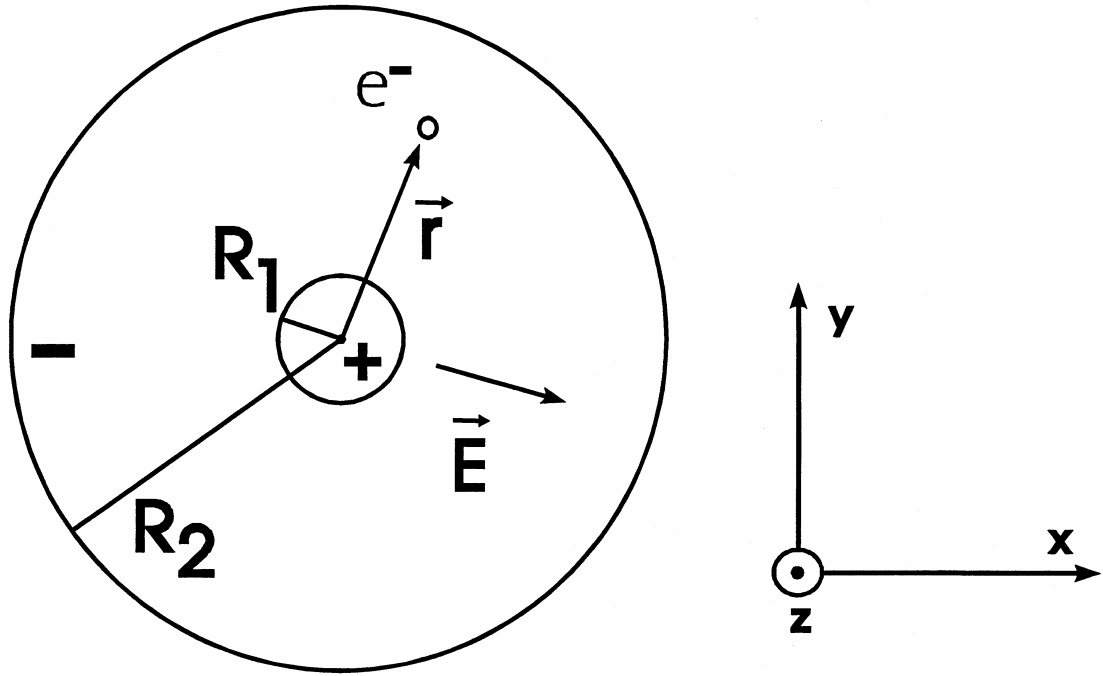

**Fig. 1.** Section in x-y plane through a cylinder capacitor with the inner and outer radii $R_1$ and $R_2$

Newton´s equation of motion for a charge q in an electric field

$$m \cdot \frac{\partial^2 \vec{r}}{\partial t^2} = q \cdot \vec{E} \qquad (5)$$

gives, after the introduction of plane polar coordinates (r,φ) and separation of the variables:

$$\left( \ddot{r} - r\dot{\varphi}^2 \right) \cdot \vec{e}_r + \left( 2\dot{r}\dot{\varphi} + r\ddot{\varphi} \right) \cdot \vec{e}_\varphi = -\frac{qk_0}{m} \cdot \frac{1}{r} \cdot \vec{e}_r \qquad (6).$$

With

$$\left(2\dot{r}\dot{\varphi}+r\ddot{\varphi}\right)=0 \qquad (7)$$

the conservation of the angular momentum in the z direction is obtained

$$L_z = m r^2 \dot{\varphi} \qquad (8)$$

as also the radial part of equation (6)

$$\left(\ddot{r}-r\dot{\varphi}^2\right)=-\frac{qk_0}{m}\cdot\frac{1}{r} \qquad (9)$$

which yields, after elimination of time, in the equation of motion

$$r\frac{d^2r}{d\varphi^2}-2\left(\frac{dr}{d\varphi}\right)^2+\frac{r^4}{a^2}-r^2=0 \qquad (10).$$

The physical constants are combined in the parameter $a^2$

$$a^2 = \frac{L_z^2}{mqk_0} \quad , \quad q\cdot k_0 > 0 \qquad (11).$$

Analytical solutions of the equation of motion (10) are up to now, unknown [1-13, 18, 19].

A first integral however, can be easily calculated as:

$$\frac{d\rho}{d\varphi}=\pm\rho^2\cdot\sqrt{A-\left(\frac{1}{\rho^2}+\ln(\rho^2)\right)} \qquad (12).$$

Here ρ is put equal to r/a. Taking $A\in\mathfrak{R}$ and A>1 for the integration constant, equation (12) possesses two real roots for each A, denoted as $\alpha_1$ and $\alpha_2$. This implies bound orbits, with the pericenter $\rho = \alpha_1$, and the apocenter $\rho = \alpha_2$. A closed integral of equation (12) is not known [14, 15, 16].

The question whether there are closed orbits within logarithmic potentials will be discussed later. Figure 2 shows a typical plot of $|d\rho/d\varphi|$.

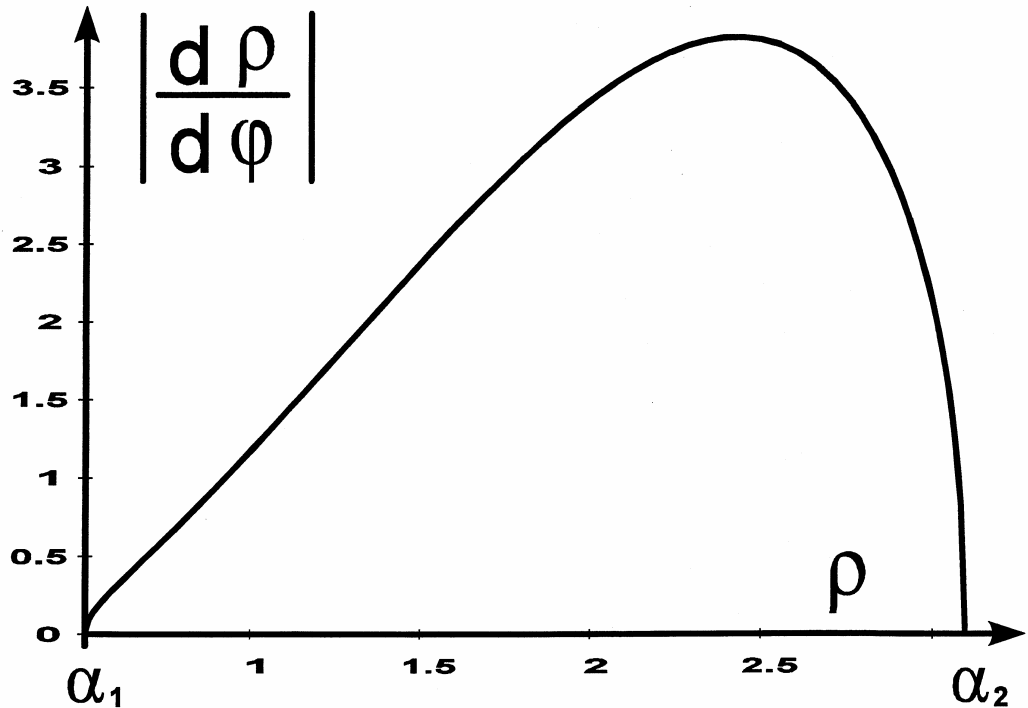

**Fig. 2.** Plot of $|d\rho/d\varphi|$ in the interval $\alpha_1 < \rho < \alpha_2$ with A=2.36495101, $\alpha_1$ = 0.5224713751 and $\alpha_2$ = 3.096694453

After substitution of a/r by z in equation (10) and collection of the differential quotients, an autonomous differential equation of the second order is obtained:

$$\frac{d^2z}{d\varphi^2} + z = \frac{1}{z} \qquad (13).$$

This differential equation can now be solved by separate power series for z and 1/z.

## 3. Solution as a Pair of Power Series

The application of power series

$$z = \sum_{k=0}^{\infty} a_k \varphi^k$$

$$\frac{1}{z} = \sum_{k=0}^{\infty} b_k \varphi^k$$

(14 a, b)

is possible, if $a_0 \neq 0$ and $b_0 \neq 0$, and if both power series have a positive radius of convergence [17]. The coefficients $a_k$ and $b_k$ are related to one another, since

$$z \cdot \frac{1}{z} = 1 \qquad (15).$$

This results in

$$a_0 \cdot b_0 = 1 \qquad (16)$$

and

$$\sum_{j=0}^{\lambda} a_j b_{\lambda-j} = 0 \quad , \quad \lambda > 0 \qquad (17).$$

The question of convergence will be dealt with separately. If equations (14a) and (14b) are substituted into equation (13)

$$b_k = a_k + (k+1)(k+2) a_{k+2} \qquad (18).$$

Here k passes through values 0, 1, 2, 3, and so on. If the existence of bound orbits is a requirement, it is possible to set $b_1 = 0$ without restriction of the general validity. On the other hand equations (17) and (18), in general, yield:

$$\sum_{j=0}^{\lambda} \left[ a_j + (j+1)(j+2) a_{j+2} \right] a_{\lambda-j} = 0 \qquad (19).$$

With (18) and (19), it becomes clear, that the functions $z(\varphi)$ and $1/z(\varphi)$ are even functions in $\varphi$, since the odd coefficients of the power series vanish:

$$a_{2k+1} = b_{2k+1} = 0 \qquad (20).$$

The coefficients $a_k$ can be found from equation (19). The coefficients $b_k$ are fixed by equation (18).

At this stage, it is convenient to express the power series (14a) and (14b) by the following equations

$$\frac{\alpha_1}{\rho} = 1 + (1-x)\sum_{k=1}^{\infty}(-1)^k p_k(x)\frac{\varphi^{2k}}{(2k)!}$$

$$\frac{\rho}{\alpha_1} = 1 + (1-x)\sum_{k=1}^{\infty}(-1)^k q_k(x)\frac{\varphi^{2k}}{(2k)!} \qquad (21\ a,\ b).$$

Here again, $\rho = r/a$, $x = \alpha_1^2$, and $\alpha_1 = b_0$. The symbols $p_k(x)$ and $q_k(x)$ denote polynomials in x, which can be determined from the recurrence relations (23a) and (23b), as follows:

- Equations (13), (21a), and (21b) yield:

$$q_k = \frac{1}{x}(p_k - p_{k+1})$$

$$p_k = 1 - x\sum_{v=1}^{k-1} q_v \qquad (22\ a,\ b).$$

Whith the product of (21a) and (21b), and taking into account the equations (22a) and (22b), the desired recurrence relations may be obtained. Their validity is restricted to values for k>1. Therefore $q_1$ and $p_1$ must be calculated separately, giving: $q_1 = -1$ and $p_1 = 1$.

There recurrence relations are:

$$q_k = Q_k - 1 + (1-x)\sum_{v=1}^{k-1}\binom{2k}{2v}\left[q_v Q_{k-v} - q_{k-v}\right]$$

$$p_k = (1+x)p_{k-1} - (1-x)\sum_{v=1}^{k-2}\binom{2(k-1)}{2v}\left[P_v + p_{k-v} - (P_v + 1)\cdot p_{k-v-1}\right]$$

(23 a, b).

For k=2 the summation in (23b) disappears. In the interests of clarity the following abbreviations have been chosen:

$$Q_k = \begin{cases} 0 & , k=1 \\ x\cdot\sum_{\mu=1}^{k-1} q_\mu & , k>1 \end{cases}$$

$$P_v = p_v - p_{v+1}$$

(24 a, b).

Tables 1 and 2 show the resulting polynomials up to k=5. Figure 3 shows the first five polynomials $q_k$. It should be noted that $0 \leq x \leq 1$.

**Table 1.** The first five polynomials $q_k(x)$.

| | | | | | |
|---|---|---|---|---|---|
| $q_1$ | -1 | | | | |
| $q_2$ | 5 | -7 x | | | |
| $q_3$ | -61 | +184 x | -127 $x^2$ | | |
| $q_4$ | 1385 | -6567 x | +9543 $x^2$ | -4369 $x^3$ | |
| $q_5$ | -50521 | +329768 x | -746910 $x^2$ | +711296 $x^3$ | -243649 $x^4$ |

**Table 2**. The first five polynomials $p_k(x)$.

| | | | | | |
|---|---|---|---|---|---|
| $p_1$ | 1 | | | | |
| $p_2$ | 1 | + x | | | |
| $p_3$ | 1 | -4 x | +7 $x^2$ | | |
| $p_4$ | 1 | +57 x | -177 $x^2$ | +127 $x^3$ | |
| $p_5$ | 1 | -1328 x | +6390 $x^2$ | -9416 $x^3$ | +4369 $x^4$ |

**Table 3.** Special values, $E_{k,0}$ are the Euler numbers in equation (28a)

| $q_k(x=0) = E_{k,0}$ | $p_k(x=0) = 1$ | $q_k(x=1) = -2^{k-1}$ | $p_k(x=1) = 2^{k-1}$ |
|---|---|---|---|

For practical applications equations (23a) and (23b) are unwieldy, particularly with a high summation index k. Furthermore, the calculation of the polynomials requires only knowledge of the coefficients, enabling a general description of the coefficients to be derived.

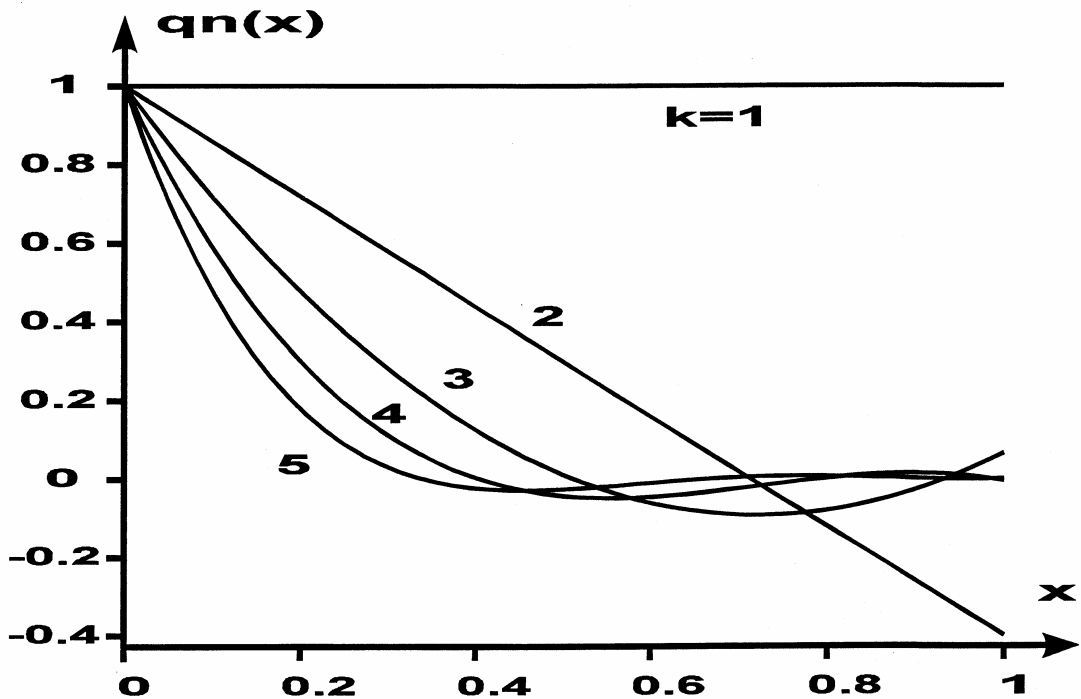

**Fig. 3.** Plot of the polynomials $qn_k(x) = q_k(x)/E_{k,0}$ in the interval $0 \leq x \leq 1$ for k = 1 to 5, standardized according to Euler numbers $E_{k,0}$, equation (28a)

## 4. General Description of the Polynomial Coefficients

A general formulation of the polynomials can be expressed by

$$q_k = \sum_{\kappa=0}^{k-1} E_{k,\kappa} x^\kappa$$

$$p_k = \sum_{\kappa=0}^{k-1} S_{k,\kappa} x^\kappa \qquad (25\ a,\ b).$$

Because of (22b), $S_{k,\kappa}$ are summations of $E_{k,\kappa}$, so that only consideration of the coefficients of the polynomials $q_k$ is required (cf. tables 1 and 2).

The substitution of (25a) into (23a) results in:

$$\sum_{\kappa=0}^{k-1} E_{k,\kappa} x^\kappa - \sum_{\kappa=0}^{k-2} T_{k,\kappa} x^{\kappa+1} + 1 =$$
$$(1-x) \sum_{\nu=1}^{k-1} \binom{2k}{2\nu} \left( \sum_{\kappa=0}^{k-3} T_{k,\nu;\kappa} x^{\kappa+1} - \sum_{\kappa=0}^{k-\nu-1} E_{k-\nu,\kappa} x^\kappa \right) \qquad (26)$$

Here abbreviations of the following form were used:

$$T_{k,\kappa} = \sum_{\lambda=1}^{k-\kappa-1} E_{\kappa+\lambda,\kappa}$$

$$T_{k,\nu;\kappa} = \sum_{\gamma=0}^{\kappa} \left( \sum_{\lambda=1}^{k-\nu-\gamma-1} E_{\lambda+\gamma,\gamma} \right) E_{\nu,\kappa-\gamma} \Bigg|_{\substack{\gamma \leq k-\nu-2 \\ \kappa-\gamma \leq \nu-1}} \qquad (27\ a,\ b).$$

If the upper index of a summation exceeds the permissible range of values, the summation has to disappear.

By ordering equation (26) in powers of x, a set of recurrence relations for $E_{k,\kappa}$ is obtained by comparison of coefficients. They are thus clearly defined, as with $\kappa \geq k$ they disappear:

$$E_{k,0} = -1 - \sum_{\nu=1}^{k-1} \binom{2k}{2\nu} \cdot E_{k-\nu,0}$$

$$E_{k,1} = T_{k,0} + \sum_{\nu=1}^{k-1} \binom{2k}{2\nu} \cdot \left(T_{k,\nu;0} - E_{k-\nu,1} + E_{k-\nu,0}\right)$$

$$E_{k,\kappa} = T_{k,\kappa-1} + \sum_{\nu=1}^{k-1} \binom{2k}{2\nu} \cdot \left(-T_{k,\nu;\kappa-2} + T_{k,\nu;\kappa-1} - E_{k-\nu,\kappa} + E_{k-\nu,\kappa-1}\right)$$

(28 a, b, c)

Equation (28a) shows the recurrence relation for the Euler numbers. Equation (28c) is valid from $\kappa = 2$ upwards.

**Table 4.** $E_{k,\kappa}$ to k=5.

| k | κ | $E_{k,\kappa}$ |
|---|---|---|
| 1 | 0 | -1 |
| 2 | 0 | 5 |
|   | 1 | -7 |
| 3 | 0 | -61 |
|   | 1 | 184 |
|   | 2 | -127 |
| 4 | 0 | 1385 |
|   | 1 | -6567 |
|   | 2 | 9543 |
|   | 3 | -4369 |
| 5 | 0 | -50521 |
|   | 1 | 329768 |
|   | 2 | -746910 |
|   | 3 | 711296 |
|   | 4 | -243649 |

With the aid of a computer algebra programme the coefficients $E_{k,\kappa}$ of the polynomials $q_k$ can be easily calculated up to any desired order.

## 5. Convergence

Proof of convergence follows from the comparison-theorem of infinite series [17]. The maximum of $|q_k(x)|$ in the range of values $0 \leq x \leq 1$ is at x=0, where

$$q_k(x=0) = E_{k,0} \qquad (29).$$

From this follows

$$\left|(-1)^k q_k(x)\right| \leq \left|E_{k,0}\right| \qquad (30).$$

The series

$$\sum c_k = \sum \left|E_{k,0}\right| \cdot \frac{\varphi^{2k}}{(2k)!} = sc(\varphi) - 1 \qquad (31)$$

possesses positive terms: $sc(\varphi) = 1/\cos(\varphi)$, $\varphi \neq \pm n\pi/2$, $n \in \mathbb{N}$. Therefore

$$\sum a_k = \sum (-1)^k q_k(x) \cdot \frac{\varphi^{2k}}{(2k)!} \qquad (32)$$

is also convergent, even absolutely convergent. The summation in (21b) is therefore convergent. As it is permissible to divide by convergent series, the summation in (21a) is also convergent [17].

## 6. Graphical Representation of the Orbits

Of special interest is the form the trajectories take in logarithmic potentials. A typical example will be considered, but first it has to be borne in mind, that if the integration constants in equation (12) are suitably chosen, closed orbits can also exist. If the angle **φ** is defined mathematically positive, and if the pericenter of the orbit is defined to be the starting point **φ**$_{Start}$ = 0 and **ρ**$_{Start}$ = **α₁**, the graphical representation of the orbits according to equation (21b) presents no problem.

A numerical integration of equation (12) gives for the angle Φ of the apocenter **ρ = α₂**, the approximation:

$$\Phi = \varphi(\alpha_2) \approx \frac{\pi}{2} \cdot \left[ 1 + \left(\sqrt{2} - 1\right) e^{\left(\frac{1-A}{2\pi}\right)} \right] \quad (33).$$

The range of values of the half period Φ lies therefore between **π/2** and **π/√2**, which means that for the first closed orbit Φ = 2π/3. The circular orbit ( d**ρ**/d**φ** = 0, **ρ** = **α₁** = **α₂** = 1, A=1 ) represents the limiting case Φ ≈ π/√2. Apart from the circular orbit, the simplest trajectory has a period of 2Φ = 4π/3. The plot of the standardized radius as a function of the angle in this case is illustrated in figure 4.

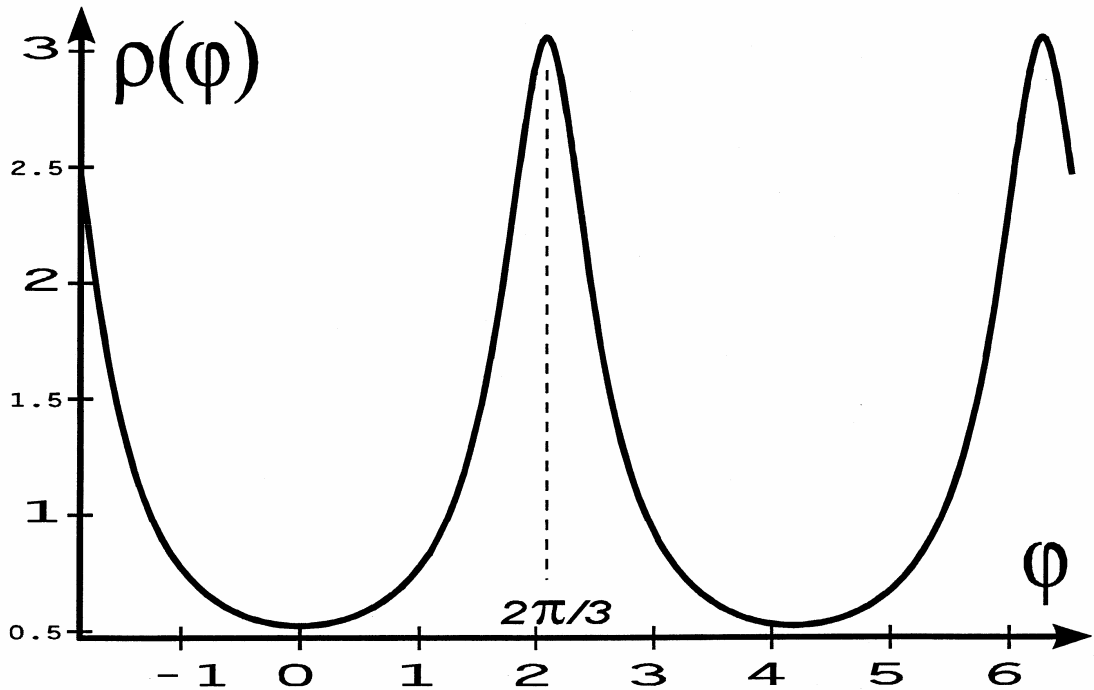

**Fig. 4.** Plot of period of **ρ(φ)** for the simplest closed trajectory. The period is 2Φ=4π/3.

Figure 5 shows the same closed orbit in plane polar coordinates. It has the shape of a rosette, similar to the three part hypocycloid. Apart from the circular orbit this is the simplest trajectory in a logarithmic potential.

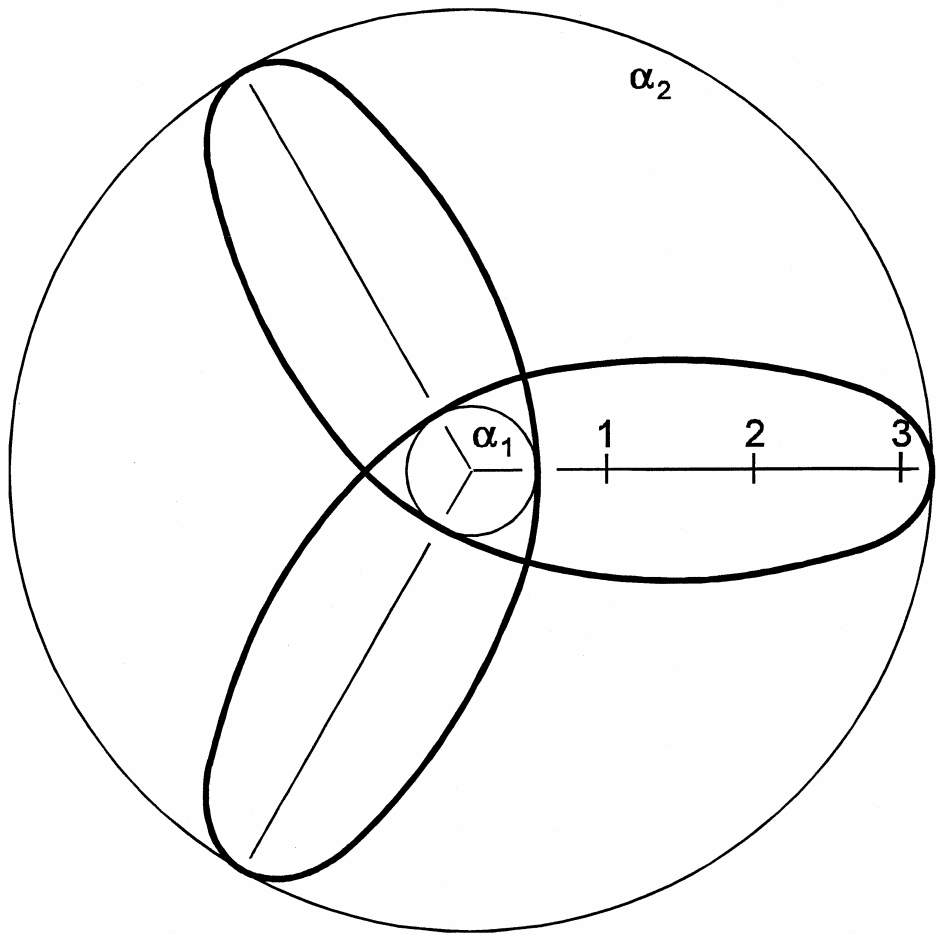

**Fig. 5.** Closed trajectory with $\Phi=2\pi/3$, the absides as well as the circles $\rho = \alpha_1$ and $\rho = \alpha_2$ are shown; A=2.36495101, $\alpha_1 = 0.5224713751$ and $\alpha_2 = 3.096694453$; see also [20]

## 7. Conclusions

The equation of motion af a point mass in a logarithmic potential leads to a quasi-harmonic autonomous second order differential equation of very simple structure. The solution of this differential equation by series ex-

- 15 -pansion of the radius vector according to the azimuth angle results in a system of coefficient polynomials.

A first integral of the equation of motion shows that by suitable choice of the integration constants bound orbits exist. Apart from the circular orbit, the trajectory with period $4\pi/3$ is the simplest closed orbit in the logarithmic potential.

The exact calculation of the orbit periods and the possible assignation of the coefficient polynomials found to a known family of polynomials, e.g. the Euler polynomials, will be a subject of further investigations.

**Contact**

Priv.-Doz. Dr. rer. nat. Erwin A.T. Wosch
Michaelstr. 2, D-52222 Stolberg, Germany
Erwin.Wosch@gmx.de
c/o  IEHK, RWTH Aachen, D-52072 Aachen, Germany
Erwin.Wosch@iehk.rwth-aachen.de